\documentclass[prc,aps,a4paper,twocolumn]{revtex4}
\usepackage{graphicx}
\usepackage{multirow}
\usepackage{hyperref}
\usepackage{amsfonts}
\usepackage{amssymb}
\usepackage{amsmath}
\usepackage{natbib}
\usepackage{dcolumn}
\usepackage{bm} 
\usepackage{epsfig}
\usepackage{pifont}
\usepackage{lmodern}
\usepackage{color,ulem}
\usepackage{hhline}
\usepackage[english]{babel}
\usepackage[autostyle]{csquotes}

\usepackage[utf8]{inputenc}

\newcommand{\bwt}{\begin{widetext}}
\newcommand{\ewt}{\end{widetext}}
\newcommand{\beq}{\begin{equation}}
\newcommand{\eeq}{\end{equation}}
\newcommand{\bea}{\begin{eqnarray}}
\newcommand{\eea}{\end{eqnarray}}

\begin{document}
\title{
Nuclear symmetry energy with mesonic cross-couplings in the effective chiral model}
\author{ Tuhin Malik$^1$}
\email{tuhin.malik@gmail.com}
\author{ Kinjal Banerjee$^1$}
\email{kinjalb@goa.bits-pilani.ac.in}
\author{ T. K. Jha$^1$}
\email{tkjha@goa.bits-pilani.ac.in} 
\author{ B. K. Agrawal$^{2,3}$}
\email{bijay.agrawal@saha.ac.in} 

\affiliation{ $^1$BITS-Pilani, Dept. of Physics, K.K. Birla Goa Campus, GOA - 403726, India.\\
$^2$Saha Institute of Nuclear physics, Kolkata 700064, India \\
$^3$Homi Bhabha National Institute, Anushakti Nagar, Mumbai -
400094, India.}

\date{\today}

\begin{abstract}
The effective chiral model is extended by introducing the
contributions from the cross-couplings between isovector and 
isoscalar mesons. These cross-couplings are found to be instrumental in
improving the density content of the nuclear symmetry energy. The nuclear
symmetry energy as well as its  slope and curvature parameters at the
saturation density are in harmony with those deduced from a diverse set
of experimental data. The equation of state for pure neutron matter at
sub-saturation densities is also in accordance with the  ones obtained
from different microscopic models. The maximum mass of neutron star
is consistent with the measurement and the radius at the canonical mass
of the neutron star is within the empirical bounds. 
\end{abstract} \maketitle

\section{Introduction}     
	Over the last decade or so there has been an extensive work and debate dedicated to 
understanding the behavior of nuclear symmetry energy theoretically as well as experimentally, 
both at low and high densities. This knowledge is helpful in understanding both finite nuclei 
and nuclear matter aspects such as Neutron Stars (NS) and supernovae dynamics, related to neutron-rich 
domain. It also helps in understanding the strong forces at the fundamental level at higher densities. 
Currently available data on nuclear masses and giant dipole polarizability have constrained 
the values of symmetry energy and its slope parameter to $J \sim 32$ MeV and $L\sim 50-80$ MeV
\cite{Trippa:2008gr,Moller:2012pxr,Tsang:2012se,Roca-Maza:2013mla,Vinas:2013hua,Roca-Maza:2015eza,Mondal:2016roo} 
at nuclear saturation density $(\rho\sim0.16~\text{fm}^{-3})$. However, little is known 
about their behavior at other densities. Motivated by this, theoretically one tries to 
modify the basic interactions so as to match with the experimental data wherever available. 
The different variants of the Relativistic Mean Field Models (RMF) could reach out to these 
values only when the contributions from the cross-coupling of $\rho$ meson to the $\sigma$ 
or $\omega$ mesons were included \cite{Todd-Rutel05,Agrawal:2010wg,Agrawal:2012rx}.  

	The models based on chiral symmetry was introduced by Gell-Mann \& Levy \cite{GellMann:1960np}. 
The importance of chiral symmetry in the study of nuclear matter was emphasized by Lee \& Wick \cite{Lee:1974ma}. 
However, the linear chiral sigma models fail to describe properties of finite nuclei. In such models, 
the normal vacuum jumps to a chirally restored  abnormal vacuum (Lee-Wick vacuum)\cite{Lee:1974ma,Lee:1974uu}. 
This phenomenon is referred to as chiral collapse problem \cite{Thomas:2004iw} and it can be overcome 
mainly in two ways. One of the approaches is to incorporate logarithmic terms of the scalar 
field in chiral potentials \cite{Furnstahl:1993wx,Heide:1993yz,Mishustin:1993ub,Papazoglou:1996hf,Papazoglou:1997uw} 
which prevents the normal vacuum from collapsing. This class of chiral models are phenomenologically 
successful in describing finite nuclei \cite{Schramm:2002xi,Tsubakihara:2006se,Tsubakihara:2007gg,Tsubakihara:2009zb}. 
However, these models explicitly break the chiral symmetry and are divergent when chiral symmetry is restored 
\cite{Furnstahl:1993wx}.

 Alternatively, the chiral collapse problem is prevented by generating the isoscalar-vector meson mass 
dynamically via Spontaneous Symmetry Breaking (SSB) by coupling the isoscalar-vector mesons with the scalar mesons 
\cite{Boguta:1983uz,Sahu:1993db}. However, the main drawback of all these models was the unrealistic 
high nuclear incompressibility ($K$). Later on, in several attempts, the higher order terms of scalar 
meson field \cite{Sahu:2000ut,Sahu:2003hz,Jha:2009kt} were introduced to ensure a reasonable $K$ 
at saturation density. The non-linear terms in the chiral Lagrangian can provide the three-body 
forces \cite{Logoteta:2015voa} which might have important roles to play at high densities. 
The effective chiral model has been used to study nuclear matter aspects such as matter at low 
density and finite temperature \cite{Sahu:2003hz}, NS structure and composition \cite{Jha:2006zi} 
and nuclear matter saturation properties. As emphasized in Ref. \cite{Sahu:2003hz}, the model parameters 
are constrained and related to the vacuum expectation value of the scalar field. Since the mass of 
the isoscalar-vector meson is dynamically generated, practically there are very few free parameters to adjust 
the saturation properties. However, this type of models had a couple of drawbacks. They yield 
the symmetry energy slope parameter, $L \sim 90$ MeV, which is a little too large. Also, the symmetry energy
at $0.1~\text{fm}^{-3}$ baryon density is $\sim 22$ MeV, which is lower than the presently estimated value 
\cite{Trippa:2008gr,RocaMaza:2012mh}. 

    In the present work, we employ the effective chiral model in which chiral symmetry breaks spontaneously.  
We extend this model by including the cross-couplings of $\sigma$ and $\omega$ mesons with the $\rho$ meson. 
We would like to see whether these terms in the interaction help in fixing the values of symmetry energy 
and its slope parameter at the saturation density. We study the effects of the cross-couplings 
on the Equation of State (EoS) for Asymmetric Nuclear Matter (ANM). 
The effects of the crustal EoS on the mass and the radius of NS
are evaluated using the method suggested recently by Zdunik \textit{et al} \cite{Zdunik:2016vza}.

      The paper is organized as follows. We briefly describe the model in Section \ref{model}.
In section \ref{results} we construct three different models with no cross-coupling, 
the $\sigma-\rho$ cross-coupling and the $\omega-\rho$ cross-coupling and corresponding results 
are discussed. Conclusions are drawn in Section \ref{con}.

\section{The Model} \label{model} 
The complete Lagrangian density for the effective chiral model which 
includes the various cross-coupling terms is given by,
\begin{eqnarray}
\label{Lag_density}    
    {\cal{L}}&=& \cal{L_0} + \cal{L_{\times}},
\end{eqnarray}
where,
\begin{eqnarray}
\label{Lag_density1}
  {\cal{L_0}}&=& \bar{\psi}_{B}\Big[\left(i \gamma_{\mu}\partial^{\mu}
                -g_{\omega}\gamma_{\mu}\omega^{\mu}-\frac{1}{2}g_{\rho}\vec{\rho_{\mu}}.
                \vec{\tau}\gamma^{\mu}\right)\nonumber\\
                &-&g_{\sigma} (\sigma + i \gamma_{5}\vec{\tau}.\vec{\pi})\Big]\psi_{B} 
                + \frac{1}{2}\left( \partial_{\mu}\vec{\pi}.\partial^{\mu}\vec{\pi}
                + \partial_{\mu}\sigma \partial^{\mu}\sigma \right)         \nonumber\\ 
               &-& \frac{\lambda}{4}\left(x^{2}-x_{0}^{2}\right)^{2}- \frac{\lambda b}
                     {6 m^{2}}(x^{2}-x_{0}^{2})^{3}  \nonumber \\ 
               &-&\frac{\lambda c}{8 m^{4}}(x^{2}-x_{0}^{2})^{4}
                    - \frac{1}{4} F_{\mu\nu}F^{\mu\nu}+ \frac{1}{2}g_{ \omega}^{2}x^{2}\left(\omega_{\mu}
                        \omega^{\mu}\right) \nonumber \\ 
                   &- &\frac{1}{4} \vec{R_{\mu \nu}}. \vec{R^{\mu \nu}} + \frac{1}{2} 
                 {m_\rho^\prime}^2\vec{\rho_{\mu}}.\vec{\rho^{\mu}} , 
\end{eqnarray}
and
\begin{eqnarray}
\label{Lag_density2}
\cal{L_{\times}} &=& \eta_1 \left(\frac{1}{2} g_\rho^2x^2 \vec{\rho_{\mu}}.\vec{\rho^{\mu}} \right)
                + \eta_2 \left(\frac{1}{2} g_{\rho}^2 \vec{\rho_{\mu}}.\vec{\rho^{\mu}}
                \omega_\mu \omega^\mu \right).
\end{eqnarray}
Here, $\psi_B$ is the nucleon isospin doublet interacting with
different mesons $\sigma$, $\omega$ and $\rho$, with the respective
coupling strengths $g_{i}$, with $i = \sigma, \omega$ and $\rho$. The $b$
and $c$ are the strength for self  couplings of scalar fields. The $\gamma^{\mu}$
are the Dirac matrices and $\tau$ are the Pauli matrices. $\cal{L_0}$ 
(Eq. (\ref{Lag_density1})) is the original Lagrangian given in Ref. \cite{Jha:2006zi}. 
Note that potential for the scalar fields $(\pi,\sigma)$ are written in
terms of a chiral invariant field $x$ given by $x^2=\pi^2 + \sigma^2$. 

$\cal{L_{\times}}$ (Eq. (\ref{Lag_density2})) is the new additional piece we
add to the original Lagrangian given in \cite{Jha:2006zi}.  It contains
cross-coupling terms between $\rho$ and $\omega$ and also between $\rho$
and $\sigma$. The coupling strength for $\sigma-\rho$ and $\omega-\rho$ 
are given by $\eta_1 g_\rho^2$ and $\eta_2 g_\rho^2$ respectively. 
The interaction of the scalar $(\sigma)$ and the pseudo-scalar $(\pi)$ 
mesons with the isoscalar-vector meson $(\omega)$ generates a dynamical mass for the $\omega$ meson
through SSB of the chiral symmetry with scalar field attaining 
the vacuum expectation value $x_0$. Then the mass of the nucleon ($m$), 
the scalar ($m_{\sigma}$) and the vector meson mass ($m_{\omega}$), 
are related to $x_0$ (vacuum expectation of $x$) through
\begin{eqnarray}
\label{ssb_mass}
m = g_{\sigma} x_0,~~ m_{\sigma} = \sqrt{2\lambda} x_0,~~
m_{\omega} = g_{\omega} x_0\ ,
\end{eqnarray}
where, $\lambda=\frac{(m_{\sigma}^2 -m_{\pi}^2)}{2 f_{\pi}^2}$ and $f_\pi
= x_0$ is the pion decay constant, which reflects the strength of SSB.
In Eq. (\ref{Lag_density2}) when $\eta_1\neq0$ there is a 
cross-interaction between  $\rho$ and $\sigma$.  Hence a fraction of $\rho$
meson mass will come from SSB. The mass of $\rho$ meson $(m_\rho)$
in this model then will be related to vacuum expectation of $x$ through
\begin{eqnarray}
\label{rho_mas}
m_\rho^2= {m_\rho^\prime}^2+ \eta_1 g_\rho^2 x_0^2 .
\end{eqnarray}
In the mean field treatment the explicit role of pion mass is ignored and 
hence $m_\pi=0$ and mesonic field is assumed to be uniform, i.e., without 
any quantum fluctuation. Then, the isoscalar-vector field $\omega$ is of 
the form $\omega_\mu=\omega_0 \delta_\mu^0 $, where $\delta_\mu^0$ is 
Kronecker delta. Note that $\omega_0$ does not depend on space-time but 
it depends on baryon density ($\rho$). 
The vector field ($\omega$), scalar field ($\sigma$) and isovector field ($\rho_3^0$) equations 
(in terms of $Y = x/x_0 = m^*/m$) are, respectively, given by:
\begin{eqnarray}
\label{field_eq1}
&& \Big[m_\omega^2 Y^2 + \eta_2 C_\rho m_\rho^2 (\rho_3^0)^2\Big] \omega_0 = g_\omega \rho, \\   
\label{field_eq2}
&&(1-Y^2)-\frac{b}{m^2 C_\omega}(1-Y^2)^2+\frac{c}{m^4 C_{\omega}^2}(1-Y^2)^3 \nonumber \\
&&+\frac{2 C_\sigma m_\omega^2 \omega_0^2}{m^2}+\frac{2 \eta_1 C_\sigma C_\rho m_\rho^2 (\rho_3^0)^2}{C_\omega m^2}
     - \frac{2 C_\sigma \rho_s}{m Y}=0 ,\\
\label{field_eq3}
&& m_\rho^2\Big[1 - \eta_1 (1-Y^2)C_\rho/C_\omega + \eta_2 C_\rho \omega_0^2\Big]\rho_3^0 \nonumber \\
&&~~~~= \frac{1}{2} g_\rho (\rho_p - \rho_n).
\end{eqnarray}
The quantity $\rho$ and $\rho_S$ are the baryon and the scalar density defined as,
\begin{eqnarray}
\label{density}
    &&\rho = \frac{\gamma}{(2 \pi)^3}\int_0^{k_F}d^3k,  \\
    &&\rho_s = \frac{\gamma}{(2 \pi)^3}\int_0^{k_F}\frac{m^*}{\sqrt{m^*{^2}+ k^2}} d^3k,    
\end{eqnarray}
where, $k_F$ is the baryon fermi momentum and $\gamma$ ( for example, $\gamma=4$ for Symmetric Nuclear 
Matter (SNM)) is the spin degeneracy factor.  $C_\sigma\equiv g_\sigma^2/m_\sigma^2$ , 
$C_\omega\equiv g_\omega^2/m_\omega^2$ and $C_\rho\equiv g_\rho^2/m_\rho^2$ are the
scalar, vector and isovector coupling parameters. The energy density $(\epsilon)$ and 
pressure $(p)$ for a given baryon density (in terms of $Y=m^*/m$) in this model is 
obtained from the stress-energy tensor, which is given as
\small{ 
 \begin{eqnarray}
\label{EoS_energy}
      \epsilon &=& \frac{1}{\pi^2}\sum_{k_n,k_p}\int_0^{k_F}k^2\sqrt{k^2+m^*{^2}}dk 
      + \frac{m^2}{8 C_{\sigma}}(1-Y^2)^2 \nonumber\\
      &&- \frac{b}{12 C_{\sigma}C_{\omega}}(1-Y^2)^3 + \frac{c}{16 m^2 C_{\sigma}
            C_{\omega}^2}(1-Y^2)^4+ \frac{1}{2}m_{\omega}^2\omega_{0}^2Y^2              \nonumber\\ 
      &&+ \frac{1}{2}m_\rho^2\Big[1- \eta_1(1-Y^2) (C_\rho/C_\omega)+ 3 \eta_2 C_\rho 
           \omega_0^2\Big] (\rho_3^0)^2, \\
\label{EoS_pressure}
   p &=& \frac{1}{3 \pi^2}\sum_{k_n,k_p}\int_0^{k_F}\frac{k^4}{\sqrt{k^2+m^*{^2}}}dk 
         - \frac{m^2}{8 C_{\sigma}}(1-Y^2)^2 \nonumber \\
     &&+ \frac{b}{12 C_{\sigma}C_{\omega}}(1-Y^2)^3 - \frac{c}{16 m^2 
            C_{\sigma}C_{\omega}^2}(1-Y^2)^4+ \frac{1}{2}m_{\omega}^2\omega_{0}^2Y^2     \nonumber\\ 
     && + \frac{1}{2}m_\rho^2 \Big[1- \eta_1(1-Y^2) (C_\rho/C_\omega)+ \eta_2 C_\rho 
              \omega_0^2\Big] (\rho_3^0)^2 .
\end{eqnarray}} \normalsize
For SNM we have to set $k_n=k_p$ and $\rho_3^0=0$. As our present knowledge of 
nuclear matter is mainly confined to normal nuclear matter density 
$(\rho_0)$, coupling constants $C_\sigma\equiv g_\sigma^2/m_\sigma^2$ and 
$C_\omega\equiv g_\omega^2/m_\omega^2$ are not free parameters in the 
Eqs.(\ref{EoS_energy},\ref{EoS_pressure}). To obtain $C_\sigma$ and $C_\omega$, we 
solve the field equations (Eqs. (\ref{field_eq1}-\ref{field_eq3})) self consistently 
while satisfying the nuclear saturation properties. Note that for different values of
$Y=x_0/x=m^*/m$, we get different values of $C_\sigma$ and $C_\omega$. 

After inclusion of cross interactions $\cal{L_{\times}}$ (Eq. (\ref{Lag_density2})) the modified 
symmetry energy $S(\rho)$ in this model is 
\small{
\begin{eqnarray}
\label{sym_e}
   S(\rho) &=& \frac{k_F^2}{6 \sqrt{k_F^2 + m*^2}}+ \frac{C_\rho k_F^3}
                   {12 \pi^2 (m^*_\rho/m_\rho)^2} + \frac{\eta_2 C_\rho^2 \omega_0^2 k_F^3 }{6 \pi^2 (m^*_\rho/m_\rho)^4 } \nonumber \\
           &&- \frac{2 \eta_2 C_\rho^2 C_\omega k_F^9}{27 \pi^6 
                   m_\omega^2 Y^4 (m^*_\rho/m_\rho)^4},
\end{eqnarray} } \normalsize
where, ${m^*}^2_\rho=m_\rho^2\Big[1- \eta_1 (1-Y^2) (C_\rho/C_\omega)+ \eta_2 C_\rho \omega_0^2\Big]$ 
and $k_F= (3 \pi^2 \rho/ 2)^{1/3}$. The coupling parameters $C_\rho$,$\eta_1$ and $\eta_2$ can 
be evaluated numerically by fixing symmetry energy $S(\rho)$ and its slope parameter $L$ at saturation density $(\rho_0)$. 
Without cross-couplings $(\eta_1=\eta_2=0)$ we revert back to the Lagrangian given in \cite{Jha:2006zi}.  

The symmetry energy can be expanded in Taylor series around saturation density$(\rho_0)$ 
as \cite{Dong:2015vga}
\bea
\label{sym_expan}
S(\rho)= J_0 + L \epsilon_1 + \frac{1}{2}K_{sym}\epsilon_1^2 +
    \frac{1}{6}Q_{sym}\epsilon_1^3 + \mathcal O(\epsilon_1^4),
\eea
where, $\epsilon_1=\frac{\rho
    - \rho_0}{3 \rho}$. The symmetry energy coefficient at $\rho_0$ is $J_0$ 
    and the other coefficient are defined at $\rho_0$
as \cite{Lattimer:2012xj},
\bea
L &=& 3 \rho \frac{\partial S(\rho)}{\partial \rho} \Big|_{\rho=\rho_0}, \\
K_{sym} &=& 9 \rho^2 \frac{\partial^2 S(\rho)}{\partial\rho^2}\Big|_{\rho=\rho_0}, \\
Q_{sym} &=& 27 \rho^3 \frac{\partial^3 S(\rho)}{\partial\rho^3}\Big|_{\rho=\rho_0}.
\eea

Similarly, the nuclear incompressibility $(K)$ of ANM can also be expanded in terms 
of $\delta$ at $\rho_0$ as $K(\delta)= K + K_{\tau} \delta^2 + \mathcal O(\delta^4)$
, where $\delta= \frac{(\rho_n-\rho_p)}{\rho}$ is the isospin asymmetry and 
$K_\tau$ is given by \cite{Chen:2009wv}
\bea
\label{ktau}
K_\tau = K_{sym} - 6 L - \frac{Q_0 L}{K},
\eea
where $Q_0= 27 \rho^3 \frac{\delta^3(\epsilon/\rho)} {\delta \rho^3}|_{\rho_0}$ in SNM .

\section{Results and Discussion} \label{results}
              As can be seen from the preceding section that the EoS of the SNM are determined by 
the coupling parameters $C_\sigma$, $C_\omega$, $b$ and $c$ (Eqs. (\ref{EoS_energy},\ref{EoS_pressure})). 
The values of these coupling parameters and resulting SNM properties at the saturation density are listed in 
Table \ref{parameter}. The values of the model parameters lie in the stable region \cite{Sahu:2009um}.
\begin{table} [!ht]  
    \caption{List of the model parameters determined from the properties of SNM 
             such as, energy per nucleon $E_0=-16~\text{MeV}$, nuclear incompressibility $K=247$ MeV
             and the nucleon effective mass $Y=m*/m=0.864$ at the saturation density $\rho_0=0.153~\text{fm}^{-3}$.
             The scalar and vector meson coupling parameters are $C_\sigma=g_\sigma^2/m_\sigma^2$ and $C_\omega=g_\omega^2/m_\omega^2$ 
             respectively. $B=b/m^2$ and $C=c/m^4$ are the parameters for the higher order self-couplings of the scalar field 
             with $m$ being the nucleon mass. The nucleon, $\omega$ meson and $\sigma$ meson masses are $939~\text{MeV}$, 
             $783~\text{MeV}$ and $469~\text{MeV}$ respectively.}
    \label{parameter}
    \centering
          \setlength{\tabcolsep}{19pt}
          \renewcommand{\arraystretch}{1.1}
    \begin{tabular}{cccc}  
        \toprule
         $C_\sigma$ & $C_\omega$ & $B$ & $C$   \\ 
         $(\text{fm}^2)$ & $(\text{fm}^2)$ & $(\text{fm}^2)$ & $(\text{fm}^4)$ \\
         \hline
          7.057 & 1.757 & -5.796 & 0.001  \\
         \toprule
    \end{tabular}
\end{table}

      The density dependence of symmetry energy $S(\rho)$ is obtained by using three different variants 
of the present model. We consider the case of no cross-coupling (NCC), the $\sigma-\rho$ cross-coupling 
(SR) and the $\omega-\rho$ cross-coupling (WR). Since the NCC model has only one free parameter 
$(\text{i.e.,}~C_\rho)$ there is not enough freedom to vary $J_0$ and $L$ independently. However, the SR and 
WR models can provide some flexibility to adjust them. Note that, in comparison to the earlier models (i.e., NCC type), 
the inclusion of cross-couplings have important implications on $S(\rho)$. The effects of the cross-couplings grow 
stronger at high densities which are relevant for the study of NS properties.
\begin{table}
   \caption{The values of the coupling constants $C_\rho,\eta_1~\text{and}~\eta_2$ are determined from various 
             symmetry energy elements. The mass of the $\rho$ meson is $770~\text{MeV}$. The values of $C_\rho$ are 
             in units of $\text{fm}^2$, $\eta_1~\text{and}~\eta_2$ are dimensionless. 
             All the symmetry energy elements are in units of MeV.}
   \centering
   \label{parameter2}
       \setlength{\tabcolsep}{8pt} 
       \renewcommand{\arraystretch}{1.1}
   \begin{tabular}{lcccc}  
       \toprule
       &  &  NCC & SR & WR \\
       \hline
Parameters &	$C_\rho$ 	 & 5.14 		& 12.28 		& 6.08          \\
   &	$\eta_1$ 	 & 0 			& -0.79 		& 0              \\
   &	$\eta_2$ 	 & 0 			&  0 			& 6.49          \\
Nuclear Matter &	$J_0$    	 & 32.5 		& 32.5 		 	& 32.5           \\
   &	$J_1$ 	         & 22.30 		& 24.49 	 	& 23.68           \\
  &	$L$ 		 & 87 			& 65 			& 65              \\
  &	$K_{sym}$ 	 & -20.09	 	& -59.16 	 	& -204.78    	 \\
  &	$Q_{sym}$ 	 & 58.73 		& 356.11 		& -88.04     	 \\
  &	$K_{\tau}$ 	 & -434	 	    	&  -368 	 	& -513    	 \\
      \toprule
   \end{tabular}
\end{table}     

	In Table \ref{parameter2} we list the values of coupling constants $(C_\rho,\eta_1~\text{and}~\eta_2)$ 
and the resulting nuclear matter properties: $J_0$, $L$, $K_{sym}$, $Q_{sym}$ and $K_\tau$ at the saturation density $\rho_0$
and $J_1$ - the symmetry energy at $\rho_1=0.1~\text{fm}^{-3}$. 
For the NCC, $C_{\rho}$ is adjusted to yield $J_0 = 32.5$ MeV. For SR(WR) model, the value of $C_{\rho}$ and 
$\eta_1(\eta_2)$ are adjusted to yield $J_0=32.5$ MeV and $L= 65$ MeV. These values are compatible with 
$J_0=31.6\pm2.66$ MeV and $L=58.9\pm16$ MeV obtained by analyzing various terrestrial experimental informations 
and astrophysical observations \cite{Li:2013ola}. It may be noted that the value of $J_{1}$ obtained for the NCC 
model shows a significant deviation from $24.1\pm 0.8$ MeV \cite{Trippa:2008gr} and $23.6\pm0.3~\text{MeV}$ 
\cite{RocaMaza:2012mh} obtained by analyzing the experimental data on isovector giant resonances, whereas, $J_1$ is 
in good agreement in case of SR and WR models. The value of $L$ obtained with NCC model is also a 
little too large. By inclusion of cross-couplings (SR and WR models) the value of $L$ is
reduced by $\sim25\%$ keeping $J_0$ fixed. In what follows, we shall present our results 
for the density dependence of symmetry energy, EoSs for the SNM and PNM and the NS properties obtained using the NCC, 
SR and WR models. We shall also compare our EoSs and the density dependence of symmetry energy with those calculated for a few 
selected RMF models, namely, NL3 \cite{Lalazissis:1996rd}, IUFSU \cite{Fattoyev:2010mx}, 
BSP \cite{Agrawal:2012rx} and BKA22 \cite{Agrawal:2010wg}. The NL3 model does not include any cross-coupling, 
the IUFSU and BSP models include the cross-coupling between $\omega$ and $\rho$ mesons, 
while, BKA22 model is obtained by including the coupling of $\rho$ mesons with the $\sigma$ mesons.  

\begin{figure} [!ht]
    \includegraphics[width=85mm]{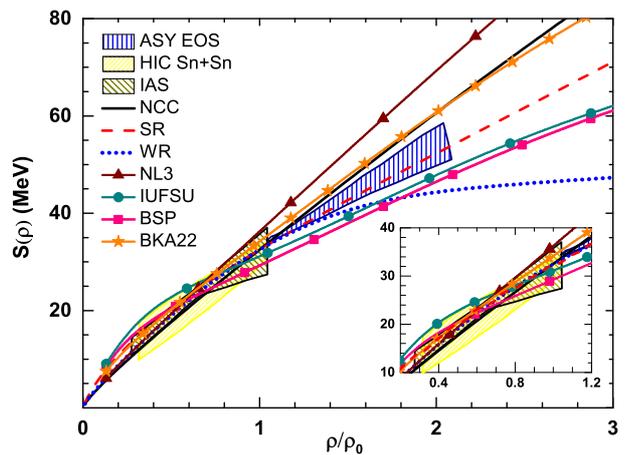}
    \caption{\label{sym_eng} \small{(Color Online) Symmetry energy as a function of scaled density $(\rho/\rho_0)$
                              is plotted for three different variants of the effective chiral model as labeled by 
                              NCC, SR and WR obtained in the present work and are compared with those for a few 
                              selected RMF models NL3, IUFSU, BSP and BKA22. The constraints on the symmetry energy 
                              from IAS \cite{Danielewicz:2013upa}, HIC Sn+Sn \cite{Tsang:2008fd}
                              and ASY-EOS experimental data \cite{Russotto:2016ucm} are also displayed.
                              The inset shows the blown up behavior of symmetry energy at low densities.}}
\end{figure}

       A lot of progress, both theoretically and experimentally, has been made to constrain symmetry 
energy at sub saturation densities. We consider the data from three important sources: simulations of low 
energy Heavy Ion Collisions (HIC) in $^{112}\text{Sn}$ and $^{124}\text{Sn}$ \cite{Tsang:2008fd}; nuclear structure studies 
by excitation energies to Isobaric Analog States (IAS) \cite{Danielewicz:2013upa} and ASY-EOS experiment 
at GSI \cite{Russotto:2016ucm}. The density dependences of the symmetry energy for NCC, SR, WR and selected RMF models
are displayed in Fig. \ref{sym_eng}. For comparison we have depicted the IAS  \cite{Danielewicz:2013upa}, 
HIC Sn+Sn  \cite{Tsang:2008fd} and ASY-EOS \cite{Russotto:2016ucm} data in the figure. It is evident that in the 
absence of any cross-couplings (NCC), the behavior of symmetry energy as a function of density is not very much 
compatible with those obtained by analyzing diverse experimental data. Remarkably the SR model satisfies all 
the above mentioned constraints. None of the considered RMF models satisfy all the symmetry energy constraints.  
The effects of various cross-couplings on the symmetry energy grow stronger at $\rho>\rho_0$. The symmetry 
energy is effectively low in WR model compared to NCC and SR models. Thus one may expect significant differences 
in the properties of NS obtained for the SR and WR models. This will be explored later in the paper.

\begin{figure} 
\includegraphics[width=85mm]{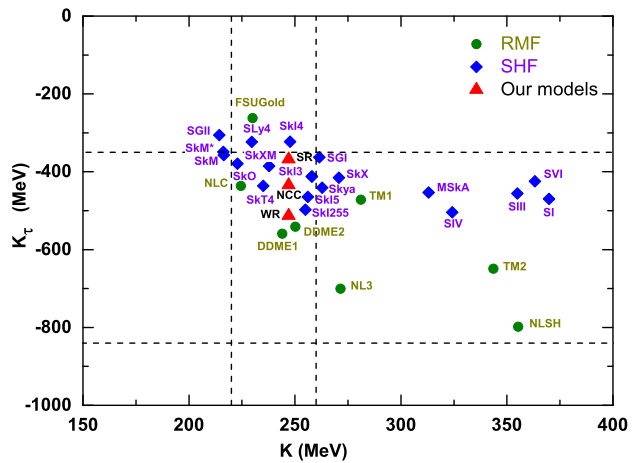}
\caption{\label{K_vs_ktau} \small{(Color Online) The values of $K$ and $K_\tau$ 
                         from different models as labeled in \cite{Sagawa:2007sp,Colo:2013yta} are compared with 
                         our models (NCC, SR and WR). The vertical and horizontal dashed lines represent the 
                         empirical ranges for $K$ and $K_\tau$ respectively.}}
\end{figure}

	 The symmetry energy elements $L$ and $K_{sym}$ predominantly determine the value of $K_\tau$ 
(Eq. (\ref{ktau})) which is required to evaluate the incompressibility of ANM. In Fig. \ref{K_vs_ktau} we compare
our values of $K_{\tau}$ with various Skyrme and RMF model predictions in $K$ vs $K_{\tau}$ plot \cite{Sagawa:2007sp}. 
The dashed lines represent the constraints on $K_\tau$ from $-840~\text{MeV}~\text{to}~-350~\text{MeV}$ 
\cite{Stone:2014wza,Pearson:2010zz,Li:2010kfa} and $K$ from $220~\text{MeV}~\text{to}~260~\text{MeV}$ 
\cite{Shlomo2006} which have been determined using various experimental data on isoscalar giant 
monopole resonances. All the three models NCC, SR and WR satisfy these bounds of $K$ and $K_\tau$.
It is to be noted that the models with a larger nuclear incompressibility $(K)$ tend to have lower $K_\tau$ value. 
As can be seen from Fig. \ref{K_vs_ktau}, several Skryme models but only three RMF models 
(NLC, DDME1 and DDME2) satisfy the bounds for $K$ and $K_{\tau}$ simultaneously. The values of $L$ for 
the nonlinear model NLC with constant coupling is $107.97$ MeV \cite{Dutra:2014qga} and that for the DDME models 
with density dependent coupling constants are $51-55$ MeV \cite{Dutra:2014qga}. The value of $L$ for NLC model 
is very large compared to presently accepted range. We have also looked into the values of $K_\tau$ and 
$K$ for the several nonlinear RMF models \cite{Alam:2016cli}. Among them a few models (BSR type) 
have $L$ between $60-70$ MeV and satisfy the constraints on $K$ and $K_\tau$. 
These models includes $\sigma-\rho$ and $\omega-\rho$ both cross-couplings.       

\begin{figure} 
\includegraphics[width=85mm]{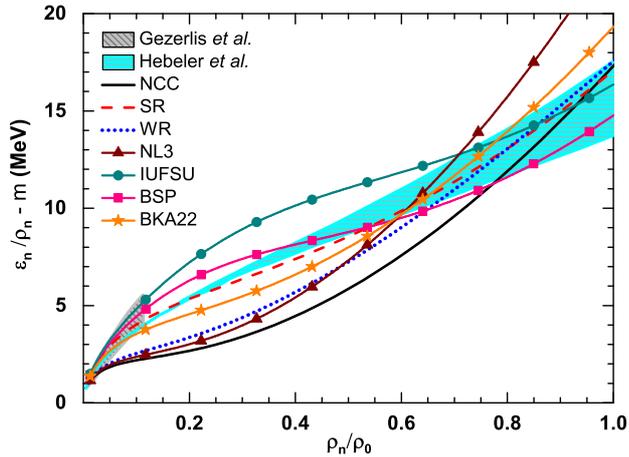}
\caption{\label{pnm} \small{(Color Online) Energy per neutron as a function of scaled neutron density 
                     $(\rho_n/\rho_0)$ for three different variants of the effective chiral model as labeled by 
                     NCC, SR and WR obtained in the present work and for a few RMF models NL3, IUFSU, BSP and BKA22
                     are compared with microscopic calculations \cite{Gezerlis:2009iw,Hebeler:2013nza} 
                     as shown by the shaded region. }}
          \end{figure}

\begin{figure} 
    \includegraphics[width=85mm]{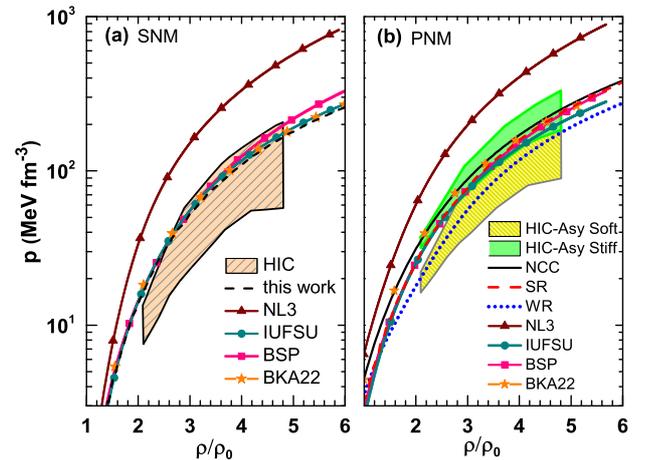} \caption{\label{HIC}
    \small{(Color Online) The pressure as a function of scaled density $(\rho/\rho_0)$ for 
                          the SNM (left) and the PNM (right). The SNM EoS for the NCC, SR and 
                          WR models are exactly the same and is labeled by \enquote{this work}.
                          For the comparison the SNM and PNM EoSs for a few RMF models NL3, IUFSU, BSP and BKA22
                          are displayed. The SNM and PNM EoSs shown by shaded regions
                          are taken from Ref. \cite{Danielewicz:2002pu} (see text for details)}} 
\end{figure}

   In Fig. \ref{pnm} we plot low density EoS for PNM for all of our three models (NCC, SR and WR). 
The low density behavior of energy per neutron for SR model is in good agreement with the results obtained 
by microscopic calculations \cite{Gezerlis:2009iw,Hebeler:2013nza} as shown by the shaded region. 
The PNM EoS for NCC and WR models do not have much overlap with the shaded region. The results for few selected 
RMF models are also displayed in the figure. Only the BSP model shows marginal overlap with the shaded region. 
In Ref. \cite{Alam:2017krb} two different families of systematically varied models with $\sigma-\rho$ and 
$\omega-\rho$ cross-couplings have been employed to study the low density behavior of asymmetric nuclear matter. 
It was found that none of the models with $\sigma-\rho$ cross-coupling satisfy the low density behavior of the PNM as 
predicted by Hebeler {\it et al} \cite{Hebeler:2013nza}. However this constraint on the PNM EoS at low 
densities are satisfied by a couple of RMF models with $\omega-\rho$ cross-coupling having $L\sim45-65$ MeV.
The EoS with the current parameterization is compared in Fig. \ref{HIC} with the experimental 
flow data obtained from the HIC \cite{Danielewicz:2002pu} for SNM and PNM EoSs. 
The later one is constructed theoretically with two extreme parameterizations, the weakest (Asy soft) 
and strongest (Asy stiff) of symmetry energy as proposed in \cite{Prakash:1988md} and as reported in 
\cite{Danielewicz:2002pu}.   
The SNM EoS is identical for all of our three models, since, the SNM properties are same. 
It is passing well through the experimental HIC data. In case of the PNM, the resulting EoSs 
for NCC and SR models pass through the upper end of HIC-Asy soft and lower end of HIC-Asy stiff, 
whereas, the PNM EoS for the WR model passes through the HIC-Asy soft only. As can be seen from 
Fig. \ref{HIC} that the influence of cross-couplings in the effective chiral model at high density
is quite strong in comparison to RMF models with similar type of cross-couplings. 
The PNM EoS for the WR model is quite softer than BSP and IUFSU at high densities. 
Similar differences can also be seen in the case of SR and BKA22 models.

      We extend our analysis to study the mass-radius relationship for static NS composed of beta equilibrated
charge neutral matter. The EoS for the core is obtained from the effective chiral model. The effects of crustal EoS 
at low densities on the mass and the radius of NS are considered in two different ways. We model the crust EoS using 
BPS EoS \cite{Baym:1971pw} in the density range $\rho\sim 4.8\times 10^{-9}~\text{fm}^{-3}~\text{to}~2.6 
\times 10^{-4}~\text{fm}^{-3}$. The crust and the core are joined using the polytropic 
form \cite{Carriere:2002bx} $p(\epsilon)=a_1+a_2 \epsilon^{\gamma}$, where the parameters $a_1$ and $a_2$ are determined in such a way that 
the EoS for the inner crust for a given $\gamma$ matches with that for the inner edge of the outer crust at one end 
and with the edge of the core at the other end. The polytropic index $\gamma$ is taken to be equal to $4/3$. 
For $\gamma=4/3$, the values of radius $R_{1.4}$ corresponding to the canonical mass of NS for the NL3 \cite{Carriere:2002bx} and 
IUFSU \cite{Piekarewicz:2014lba} RMF models are with in $\sim2\%$ in comparison to those obtained by treating the inner crust 
in the Thomas Fermi approach \cite{Grill:2014aea}. Alternatively, we estimate the contributions of the crust EoS to 
the NS radius and mass using the core crust approximation approach given in \cite{Zdunik:2016vza} referred hereafter 
ZFH method. This method enables one to estimate total mass and radius of a NS including the crust contributions 
very accurately for NS mass larger than $1~\text{M}_\odot$. In the ZFH method the radius and the mass of NS are given by       
\begin{eqnarray}
\label{r_haensel}
        R &=& \frac{R_{core}}{1- (\alpha - 1 ) (R_{core} c^2/2 G M - 1)}, \\
\label{m_haensel}        
        M &=& M_{crust}+ M_{core},
\end{eqnarray}        
with, 
\begin{eqnarray}
\label{mcrust_haensel}
       M_{crust} &=& \frac{4 \pi P_{cc} R_{core}^4}{G M_{core}}(1 - \frac{2 G M_{core}}{R_{core} c^2}).
\end{eqnarray}
In the above equations $\alpha=(\mu_{cc}/\mu_0)^2$, $\mu_{cc}~\text{and}~\mu_0$ are the chemical potential at 
transition density $(\rho_{cc})$ and at neutron star surface respectively. $R_{core}$ and $M_{core}$ are the 
radius and mass of NS core. $P_{cc}$ is pressure at transition density. The transition density $(\rho_{cc})$ 
is mostly in the range $0.4$ to $0.6~\rho_0$ for $L$ typically ranging from $30$ to $120~\text{MeV}$ 
\cite{Ducoin:2011fy}. In the present work we have taken $\rho_{cc}/\rho_0= 0.3,0.4~\text{and}~0.5$.  
               
	 Comparison of the results of the two approaches is given in Table \ref{ToV_result}.
The maximum mass of the NS is sensitive neither to the methods used to estimate the crust effects 
nor to the choice of transition density. The WR model, which includes $\omega-\rho$ 
cross-coupling, does not satisfy the maximum mass constraint as imposed by 
PSR $J0348+0432$ $(M = 2.01\pm0.04~M_\odot)$ \cite{Antoniadis:2013pzd}. This disfavors the WR model. 
The values of $R_{1.4}$ obtained using BPS EoS for the outer crust and polytropic EoS for the inner 
crust are little too large compared to those for the ZFH method. We find that by including $\sigma-\rho$ 
coupling (SR) $R_{1.4}$ are smaller compare to the NCC model which does not include any cross-coupling term. 
The radius of NS is sensitive to transition density. Using the strong correlation between transition 
density $(\rho_{cc})$ and $L$, we found the values of $\rho_{cc}$ to be $0.061~\text{fm}^{-3}~(\sim0.4~\rho_0)$ 
for NCC and $0.077~\text{fm}^{-3}~(\sim0.5~\rho_0)$ for SR and WR models respectively \cite{Grill:2014aea}.  
\begin{table}  
\label{ToV_result}
\caption{The maximum mass and radius of NS composed of $\beta-$ equilibrated matter are listed.
        The total mass and radii following the ZFH method are obtained by using Eqs. \ref{r_haensel}-\ref{mcrust_haensel}. 
        These are compared with the ones calculated from the BPS and polytropic EoSs for the 
        outer and inner crusts, respectively.
        $\rho_{cc}/\rho_0$ is the scaled transition density. 
         $M_{max}$ , $R_{max}$ and  $R_{1.4}$ are the NS maximum mass, radius at maximum mass and the radius at 
         $1.4~M_\odot$ respectively.}
      \setlength{\tabcolsep}{3.2pt}
      \renewcommand{\arraystretch}{1}
\begin{tabular}{ >{\centering}m{0.6cm} *7c }
\toprule
\multirow{3}{*}{$\frac{\rho_{cc}}{\rho_0}$} & \multirow{3}{*}{Model} & \multicolumn{3}{c}{BPS+polytropic EoS} & \multicolumn{3}{c}{ZFH method} \\ 
 &  & $M_{max}$ & $R_{max}$ & $R_{1.4}$ & $M_{max}$ & $R_{max}$ & $R_{1.4}$ \\ 
 & & {\tiny$M_\odot$} & {\tiny km} & {\tiny km} & {\tiny $M_\odot$} &  {\tiny km} & {\tiny km}  \\ \hline
\multirow{4}{*}{0.3} & NCC 	& 1.97   & 11.55   & 13.31    & 1.97   & 11.48   & 13.12 \\
                     & SR   & 1.97   & 11.24   & 12.75    & 1.97   & 11.20   & 12.71 \\
                     & WR   & 1.84   & 10.74   & 12.22    & 1.84   & 10.67   & 12.03  \\ \hline
\multirow{4}{*}{0.4} & NCC 	& 1.97   & 11.64   & 13.57    & 1.97   & 11.48   & 13.12  \\
		             & SR  	& 1.97   & 11.28   & 12.87    & 1.97   & 11.21   & 12.72  \\ 
                     & WR 	& 1.84   & 10.83   & 12.41    & 1.84   & 10.67   & 12.03  \\ \hline
\multirow{4}{*}{0.5} & NCC 	& 1.97   & 11.77   & 13.90    & 1.97   & 11.50   & 13.13   \\
		             & SR   & 1.97   & 11.35   & 13.04    & 1.97   & 11.24   & 12.72   \\
                     & WR  	& 1.84   & 10.92   & 12.62    & 1.84   & 10.67   & 12.03   \\ \toprule
\end{tabular}
\end{table}
\begin{figure} 
\includegraphics[width=85mm]{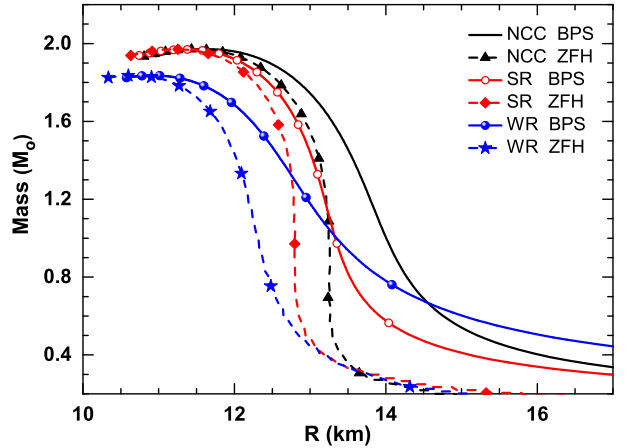}
\caption{\label{mr_curve} \small{(Color Online) The mass-radius relationships for the NCC, SR and WR models are displayed. 
                                 The effects of the crustal EoSs are incorporated by using explicitly the BPS and
                                 polytropic EoSs (solid lines) at low densities and alternatively using
                                 the ZFH method (dashed lines).}}
\end{figure}
The mass radius relationship for the NS for all of our 
three models obtained using respective values of the transition densities are plotted in Fig. \ref{mr_curve}. 
The dashed lines are obtained using the ZFH method in which the effects of the crust EoS were approximated and 
the solid lines are obtained using BPS and the polytropic EoSs for the outer and the inner crust respectively. 
It is found that the value of $R_{1.4}$ is decreased by $\sim0.5$ km in SR model compared to NCC model. The $R_{1.4}$ 
of SR is consistent with $11.9\pm1.22$ km ($90\%$ confidence) obtained by constraining symmetry energy at 
saturation density from various experimental information and theory \cite{Lattimer:2012xj}. The NS maximum mass
$\text{M}_{\rm max}=$ $2.79$, $1.94$, $2.02$, $2.04~\text{M}_\odot$ and the radius $R_{1.4}=$ $14.66$, $12.49$, $12.64$, $13.28$ 
km for the selected RMF models NL3, IUFSU, BSP and BKA22 respectively. The RMF models
such as IUFSU and BSP with $\omega-\rho$ cross-coupling readily yield $\text{M}_{\rm max}\sim2~\text{M}_\odot$, since, 
the softening of the EoS due to the inclusion of this cross-coupling is not as strong as in the case of effective chiral model. 

Results obtained for the SR model can be summarized in the following way. 
It yields symmetry energy $J_0=32.5$ MeV, symmetry energy slope parameter $L=65$ MeV, nuclear incompressibility 
$K=247$ MeV and the asymmetry term of nuclear incompressibility $K_\tau=-368$ MeV at saturation density 
$\rho_0=0.153$ fm$^{-3}$. It also yields symmetry energy $J_1=24.49$ MeV at density $0.1$ fm$^{-3}$, NS maximum 
mass $1.97~M_\odot$ and radius $R_{1.4}=12.72$ km. All these values are within presently accepted range. 
The SR model also satisfies all the discussed constraints from microscopic calculations for low density PNM EoS, 
density dependence of symmetry energy, HIC data for SNM EoS and HIC-Asy stiff data for PNM EoS.

The contributions of the exotic degrees of freedom, such as hyperons, kaons etc. to the properties of NS
are not considered in the present work. In general, the presence of strange particles softens the EoS
and reduce the NS maximum mass. In particular, the inclusion of hyperons in the effective chiral model 
(i.e. NCC type) tend to reduce the NS maximum mass by $\sim0.3~\text{M}_\odot$ \cite{Jha:2006zi}. 
The influence of hyperons on the NS properties, however, are very sensitive to the choice of the 
meson-hyperon couplings. It has been shown that sizable fraction of hyperons may exist in the NS 
with a mass $2~\text{M}_\odot$ , provided, strong repulsive hyperon-hyperon interaction is introduced 
through strange $\phi$ mesons \cite{Weissenborn:2011kb,Sulaksono:2012ny,Bizarro:2015wxa}.

\section{Conclusion} \label{con}
       We have extended the effective chiral model by including the contributions from $\sigma-\rho$ and 
$\omega-\rho$ cross-couplings. The inclusion of cross-couplings involving $\rho$ meson has helped to improve overall 
behavior of the density dependence of the symmetry energy. 

       We have discussed three different variants of effective chiral model in this paper. The model with no cross-coupling (NCC), $\sigma-\rho$ 
cross-coupling (SR) and $\omega-\rho$ cross-coupling (WR). NCC model yields the value of symmetry energy slope parameter 
$(L=87~\text{MeV})$ which is a little too large and symmetry energy at crossing density $0.1~\text{fm}^{-3}$ $(J_1=22.3~\text{MeV})$ 
which is low compared to presently estimated values. The low-density behavior of PNM EoS for both NCC and WR models does 
not match well with the range of values proposed by microscopic calculations \cite{Gezerlis:2009iw,Hebeler:2013nza}. 
The WR model gives NS maximum mass to be $1.86~M_\odot$ which is very less compare to the mass observed for the 
PSR $J0348+0432$ $(M = 2.01\pm0.04~M_{\odot})$ \cite{Antoniadis:2013pzd}. 

      For the SR model, the overall behavior of the density dependence of the symmetry energy agree well with IAS, HIC Sn+Sn 
and ASY-EOS data in comparison to those for the NCC and WR models. The symmetry energy at the saturation density and at the crossing 
density ($\rho_1 = 0.1\text{ fm}^{-3}$) are in harmony with the available empirical data. The value of the symmetry energy slope 
and the curvature parameters are in accordance with those deduced from the diverse set of experimental data for the finite nuclei. 
The pure neutron matter EoS at sub-saturation densities  passes well through the range of values suggested by the microscopic 
models \cite{Gezerlis:2009iw,Hebeler:2013nza}. The NS maximum mass is $1.97~M_\odot$ which is consistent with the observational constraint. 
The value of $R_{1.4}$ is within the empirical bounds. The SR model satisfies all the discussed constraints which suggest that   
the inclusion of $\sigma-\rho$ cross-coupling in the effective chiral model is indispensable. We have also compared our results 
with a few selected RMF models. In general, it is found that the effects of various cross-couplings within the RMF models
are weaker compare to those in the effective chiral model. This effects are more prominent for the models with $\omega-\rho$
cross-coupling.

\begin{acknowledgements}
The authors TM \& TKJ would like to thank DAE-BRNS for the support (Ref: 2013/37P/5/BRNS).
TM would also like to thank SINP for the hospitality provided during his visit for this work. 
TM would like to thank Chiranjib Mondal and Naosad Alam for useful comments.
\end{acknowledgements}


\end{document}